\documentclass[prd,twocolumn,superscriptaddress,amsmath,%
amssymb,nofootinbib]{revtex4-1}

\begin{document}


\title{(Super-)renormalizably dressed black holes}

\author{Eloy Ay\'on-Beato}
\email{ayon-beato-at-fis.cinvestav.mx}
\affiliation{Departamento de F\'{\i}sica, CINVESTAV--IPN, Apdo.
Postal 14--740, 07000 M\'exico~D.F., M\'exico}
\affiliation{Instituto de Ciencias F\'isicas y Matem\'aticas,
Universidad Austral de Chile, Casilla 567 Valdivia, Chile}

\author{Mokhtar~Hassa\"ine}
\email{hassaine-at-inst-mat.utalca.cl}
\affiliation{Instituto de Matem\'atica y F\'isica, Universidad de Talca,
Casilla 747 Talca, Chile}

\author{Julio A.\ M\'endez-Zavaleta}
\email{jmendezz-at-fis.cinvestav.mx}
\affiliation{Departamento de F\'{\i}sica, CINVESTAV--IPN, Apdo.
Postal 14--740, 07000 M\'exico~D.F., M\'exico}
\affiliation{Instituto de Ciencias F\'isicas y Matem\'aticas,
Universidad Austral de Chile, Casilla 567 Valdivia, Chile}

\begin{abstract}
Black holes supported by self-interacting conformal scalar
fields can be considered as renormalizably dressed since the
conformal potential is nothing but the top power-counting
renormalizable self-interaction in the relevant dimension. On
the other hand, potentials defined by powers which are lower
than the conformal one are also phenomenologically relevant
since they are in fact super-renormalizable. In this work we
provide a new map that allows to build black holes dressed with
all the (super-)renormalizable contributions starting from
known conformal seeds. We explicitly construct several new
examples of these solutions in dimensions $D=3$ and $D=4$,
including not only stationary configurations but also
time-dependent ones.
\end{abstract}

\maketitle

\section{\label{sec:intro}Introduction}

The nontrivial interaction between matter and geometry
mathematically encoded through the equations proposed by
Einstein a century ago is one of the crucial paradigms of
General Relativity. As it is well-known, the nonlinearity of
Einstein equations makes the task of finding physically
interesting solutions very hard. In order to render this
problem more easy to handle, the presence of certain symmetry
can be primordial to simplify the problem or as a tool to
generate nontrivial solutions from more simpler ones.

As a concrete example, we mention the case of a
self-gravitating scalar field conformally coupled to gravity.
Indeed, in this case, the conformal symmetry of the matter
source implies that the scalar curvature is zero which in turn
considerably simplifies the field equations for a simple
ansatz, yielding to the so-called Bekenstein black hole
solution \cite{Bekenstein:1974sf,BBMB}. This solution was in
fact discovered by Bekenstein exploiting the conformal
machinery enjoyed by the system by mapping the original action
to the one of a minimally coupled scalar field. The inclusion
of an electromagnetic field, also conformally invariant in four
dimensions, is straightforward and yields to the charged
version of the Bekenstein black hole. One may be tempted to
conjecture that any conformal contribution added to the
original problem will provide a generalization of this
solution. However, as shown in \cite{AyonBeato:2002cm}, the
inclusion of the conformally invariant self-interaction turns
to be incompatible with the asymptotic flatness of the black
hole. As will be discussed later, the prize to be paid in order
to consider this possibility is to add a cosmological constant,
which clearly modifies the asymptotic behavior.

The mere existence of the Bekenstein black hole, together with
its very particular properties has attracted a lot of attention
as it can be seen by reading the current literature on the
topic. For example, this solution provided one of the first
counterexamples to the old statements of the ``no-hair''
conjecture, and its uniqueness has been established in
\cite{XZ}. Nevertheless, the solution suffers from a
pathological behavior due to the fact that the scalar field diverges
at the horizon which makes rather obscure its physical
interpretation \cite{Sudarsky:1997te}. A way of circumventing
this problem is to reconsider the inclusion of the
self-interaction that does not spoil the conformal invariance
but this time together with a cosmological constant, whose
effect is precisely to push this singularity behind the
horizon. The resulting configuration is the so-called
Mart\'{\i}nez-Troncoso-Zanelli (MTZ) black hole
\cite{Martinez:2002ru}, which also allows a charged
generalization \cite{Martinez:2005di}. Due to the presence of
the cosmological constant, the existence of these
configurations is ensured not only for horizons with spherical
topology but also with hyperbolic one. In
Ref.~\cite{Caldarelli:2013gqa}, it has been shown that the
inclusion of axion fields coupled to the scalar field allows
the existence of black hole configurations with planar horizon
topology, valid even for more general nonminimal couplings than
the conformal one.

Black holes dressed by a conformal scalar field in presence of
a cosmological constant were in fact first analyzed in $2+1$
dimensions by Mart\'{\i}nez and Zanelli (MZ)
\cite{Martinez:1996gn}. The MZ black hole was later generalized
to its conformally self-interacting version
\cite{Henneaux:2002wm}, and a further nonlinearly charged
extension preserving the conformal invariance of the full
source has been given recently in \cite{Cardenas:2014kaa}. It
is even known the exact gravitational collapse to the neutral
self-interacting version \cite{Xu:2014xqa}.

In contrast, for higher dimensions ($D>4$) the situation is
quite different; the Bekenstein conformal mapping can be
lifted, but it produces only naked singularities
\cite{Xanthopoulos:1992fm}. This result can be proved in full
generality by straightforwardly integrating the static and
spherically symmetric equations of motion, which in turn
implies the uniqueness of the Bekenstein black hole within all
the asymptotically flat black hole solutions allowed by a
conformal scalar field in any dimension \cite{Klimcik:1993ci}.
No similar result in such generality is known in presence of a
cosmological constant. Nevertheless, it can be shown that in
the case of the simplest metric ansatz involving a single
function, the only black hole solutions are those previously
mentioned for $D=3$ and $D=4$
\cite{AyonBeato:2001sb,Martinez:2009kua}. For completeness, we
also mention that generalizing the conformal action in higher
dimensions via nonminimal couplings of the scalar field with
higher-curvature Lovelock terms built out of a Weyl connection
(by construction conformally invariant) whose vectorial part is
a pure-gauge contribution of the derivatives of the scalar
field, asymptotically (A)dS black hole solutions can be
obtained \cite{Giribet:2014bva}. Additionally, generalizing
also the gravity action to the Lovelock one it is possible to
find topological black hole configurations for a
self-interacting scalar field with standard nonminimal coupling
to gravity \cite{LovelockBHSF}; some of those examples include
conformally coupled scalar fields.

Another important aspect of the conformal symmetry as guiding
principle is the fact that the power defining the conformal
self-interaction in any dimension $D$ is just $2D/(D-2)$. In
mass units $c=1=\hbar$, this implies that the conformal
coupling constant is dimensionless and consequently these
theories are power-counting renormalizable
\cite{Peskin:1995ev}. From the phenomenological point of view,
self-interactions with lower powers than the conformal one are
also relevant, since its coupling constants have positive mass
dimension and they define super-renormalizable theories
\cite{Peskin:1995ev}. In fact, the mechanism explaining the
spontaneous symmetry breaking of gauge theories is naturally
modeled with these kind of contributions
\cite{Englert:1964et,Higgs:1964pj}. Due to the recent
experimental confirmation of the existence of scalar bosons in
nature \cite{Aad:2012tfa,Chatrchyan:2012ufa}, exploring their
self-gravitating behavior is an important task to study, even
in their more extremal realization, i.e.\ their possible
gravitational collapse to black holes. Examples of the final
state of this collapse are given by the (super-)renormalizably
dressed black holes found by Anabalon and Cisterna (AC)
\cite{Anabalon:2012tu}, for which the scalar field is no longer
conformally invariant but retains its conformal coupling.
Indeed, conformal invariance is explicitly broken in their
source because the potential involves additional powers of the
scalar field, lower than the conformal one. One of the aim of
the present paper is precisely to explain the emergence of
these additional contributions to the potential by means of a
generating tool as the ones emphasized at the beginning. This
tool does not only explain the existence of the AC black hole,
but also gives rise to many new scalar black holes supported
with (super-)renormalizable self-interactions.

In this work we concretely show that the Einstein equations
with cosmological constant for a self-interacting conformally
invariant scalar field can be mapped to their counterpart
having as source an also conformally coupled scalar field, but
with the difference of being subject to a more general
self-interaction that explicitly breaks conformal invariance.
We prove this result by showing that, via a very precise map,
the corresponding actions are proportional to each other. This
mapping is realized through the composition of a shift on the
original scalar field and a conformal transformation acting on
the metric and the scalar field. For dimensions $D=3, 4$ and
$6$ which correspond to the dimensions where the conformal
power $2D/(D-2)$ is an integer, the expression of the resulting
potential involves all the integer powers of the scalar field
until the conformal one yielding to a (super-)renormalizable
self-interaction. Consequently with the fact that these
self-interactions involve in general more coupling constants
than the conformal one, this is a one-way map allowing to
obtain self-gravitating solutions of a scalar field conformally
coupled to gravity and self-interacting via
(super-)renormalizable contributions from any self-gravitating
conformal seed, but the converse is not true.

The plan of the paper is the following. In the next section, we
present the new mapping from a self-gravitating conformal
scalar field to another self-gravitating one which is
(super-)renormalizably self-interacting and conformally coupled
to gravity, both in presence of its respective cosmological
constants. In dimensions $D=4$ and $D=3$, where black hole
solutions dressed by a conformal scalar field are known, we use
this mapping to explicitly exhibit new solutions of a
(super-)renormalizable and conformally coupled scalar source in
Secs.~\ref{sec:D=4} and \ref{sec:D=2+1}, respectively. Finally,
the achieved conclusions are given in Sec.~\ref{sec:conclu}.

\section{\label{sec:genconf}From renormalizable to
(super-)renormalizable sources}

Our starting point is to consider the $D$-dimensional action of
a self-gravitating conformal scalar field in presence of a
cosmological constant
\begin{align}\label{eq:Sconformal}
S[g,\Phi]={}&\int d^{D}x\sqrt{-g} \biggl(\frac{R-2\Lambda}{2\kappa}
-\frac{1}{2}\partial_{\mu}\Phi\partial^{\mu}\Phi\nonumber\\
&\qquad\qquad\qquad-\frac{1}{2}\xi_{D}R\Phi^{2}
-\lambda\Phi^{\frac{2D}{D-2}}\biggr).
\end{align}
Here, $R$ stands for the scalar curvature, $\Lambda$ represents
the cosmological constant, $\lambda$ is the coupling constant
of the conformal potential and the conformal coupling is given
by
\begin{equation}
\xi_{D}=\frac{D-2}{4(D-1)}.
\end{equation}
This is the precise value of the nonminimal coupling to gravity
that ensures the scalar contribution to the action
(\ref{eq:Sconformal}) to be invariant, up to a boundary term,
under a conformal transformation
\begin{equation}\label{eq:conftransf}
g_{\mu\nu}\mapsto\Omega^2\,g_{\mu\nu},\qquad
\Phi\mapsto\Omega^{\frac{2-D}{2}}\,\Phi,
\end{equation}
where $\Omega$ is an arbitrary local function.

The field equations for the conformal scalar field arising from
the variation of the action (\ref{eq:Sconformal}) read
\begin{subequations}\label{eq:NMGSconf1}
\begin{align}
G_{\mu\nu}+\Lambda{g}_{\mu\nu} &= {\kappa}T_{\mu\nu},
\label{eqmotion1}\\
\Box\Phi-{\xi_D}R\Phi &= \frac{2\lambda D}{D-2}\Phi^{\frac{D+2}{D-2}},
\end{align}
where the conformally invariant energy-momentum tensor is
defined by
\begin{align}
T_{\mu\nu}={}&\nabla_{\mu}\Phi\nabla_{\nu}\Phi
-g_{\mu\nu}\left(\frac{1}{2}\nabla_{\sigma}\Phi\nabla^{\sigma}\Phi
+\lambda\Phi^{\frac{2D}{D-2}}\right)\nonumber\\
&+\xi_D( g_{\mu\nu}\Box - \nabla_{\mu}\nabla_{\nu}+G_{\mu\nu} )\Phi^2.
\end{align}
\end{subequations}

Here, we show that a slightly generalization of the
transformations (\ref{eq:conftransf}) can also induce some
interesting features not only for the matter source but for the
full action (\ref{eq:Sconformal}). Indeed, we prove that there
exists a special conformal frame, defined by taking a precise
power of the conformal factor as an affine function of the
scalar field, where a simple shift of the scalar field permits
to map the action (\ref{eq:Sconformal}) into a similar action
but with a more involved self-interaction potential. More
precisely, under the following one-parameter mapping defined by
\begin{subequations}\label{eq:map}
\begin{align}
\bar{g}_{\mu\nu}&=\left(a\sqrt{\kappa\xi_{D}}\Phi+1\right)^{\frac{4}{D-2}}g_{\mu\nu},
\label{eq:mapg}\\
\bar{\Phi}&=\frac{1}{\sqrt{\kappa\xi_{D}}}\frac{\sqrt{\kappa\xi_{D}}\Phi+a}
{a\sqrt{\kappa\xi_{D}}\Phi+1},
\label{eq:mapPhi}
\end{align}
\end{subequations}
the action for a self-gravitating conformal scalar field
(\ref{eq:Sconformal}) transforms to a new action
\begin{equation}\label{eq:mapaction}
\bar{S}[\bar{g},\bar{\Phi}]=(1-a^{2})\,S[g,\Phi],
\end{equation}
which also describes a self-gravitating conformally coupled
scalar field
\begin{subequations}\label{eq:Sbar}
\begin{align}
\bar{S}[\bar{g},\bar{\Phi}]={}&\int
d^{D}x\sqrt{-\bar{g}}\biggl(\frac{\bar{R}-2\bar{\Lambda}}{2\kappa}
-\frac{1}{2}\partial_{\mu}\bar{\Phi}\partial^{\mu}\bar{\Phi} \nonumber\\
&\qquad\qquad\qquad-\frac{1}{2}\xi_{D}\bar{R}\bar{\Phi}^{2}-V(\bar{\Phi})\biggr),
\end{align}
but subject to a different self-interaction potential given by
\begin{align}
V(\bar{\Phi})={}&\frac{1}{\left(1-a^{2}\right)^{\frac{D+2}{D-2}}}
\Bigg\{\frac{\Lambda}{\kappa}
\left[\left(1-a\sqrt{\kappa\xi_{D}}\bar{\Phi}\right)^{\frac{2D}{D-2}}-1\right]\nonumber\\
&+\lambda\left[ \left(\bar{\Phi}-\frac{a}{\sqrt{\kappa\xi_{D}}}\right)^{\frac{2D}{D-2}}
-\left(-\frac{a}{\sqrt{\kappa\xi_D}} \right)^{\frac{2D}{D-2}}\right]\Bigg\},
\label{eq:potential}
\end{align}
and in presence of a modified cosmological constant
\begin{equation}\label{eq:barLambda}
\bar{\Lambda}=\frac{\kappa}{\left(1-a^{2}\right)^{\frac{D+2}{D-2}}}
\left[\frac{\Lambda}{\kappa}+
\lambda\left(-\frac{a}{\sqrt{\kappa\xi_{D}}}\right)^{\frac{2D}{D-2}}\right].
\end{equation}
\end{subequations}
Notice that these precise definitions ensure that the new potential
does not contain zeroth-order terms, and hence the modified
cosmological constant lacks of further contributions.

We are entitled to ask the reasons for which this mapping can
be relevant. Apart from generating new solutions from conformal
ones, the mapping has also a nice feature in dimensions $D=3,
4$ and $6$ where the conformal power $2D/(D-2)$ is an integer.
Indeed, in those cases, the resulting self-interaction
(\ref{eq:potential}) enhances the original conformal one with
all the power-counting super-renormalizable contributions,
i.e.\ it becomes a polynomial of degree $2D/(D-2)$,
\begin{equation}\label{eq:SuperRenorm}
V(\bar{\Phi})=\lambda_1\bar{\Phi}+\lambda_2\bar{\Phi}^2+\cdots+
\lambda_{2D/(D-2)}\bar{\Phi}^{2D/(D-2)}.
\end{equation}

Since this mapping is operated at the level of the actions and
only change them by a global multiplicative factor, the
solutions of the field equations (\ref{eq:NMGSconf1}) can be
mapped to solutions of the field equations arising from the
variation of the action (\ref{eq:Sbar}) which are given by
\begin{subequations}\label{eq:FE}
\begin{align}
\bar{G}_{\mu\nu}+\bar{\Lambda}\bar{{g}}_{\mu\nu} &= {\kappa}{\bar T}_{\mu\nu},
\label{eq:motion2}\\
\bar{\Box}\bar{\Phi}-{\xi_D}\bar{R}\bar{\Phi} &= \frac{dV(\bar{\Phi})}{d\bar{\Phi}},
\label{eq:motion3}
\end{align}
where now the new energy-momentum tensor is defined by
\begin{align}
\bar{T}_{\mu\nu}={}&\partial_{\mu}\bar{\Phi}\partial_{\nu}\bar{\Phi}
-\bar{g}_{\mu\nu}\left(\frac{1}{2}\partial_{\sigma}\bar{\Phi}\partial^{\sigma}\bar{\Phi}
+{V}(\bar{\Phi})\right)\nonumber\\
&+\xi_D( \bar{g}_{\mu\nu}\bar{\Box}
-\bar{\nabla}_{\mu}\bar{\nabla}_{\nu}+\bar{G}_{\mu\nu} )\bar{\Phi}^2,
\end{align}
\end{subequations}
and involves a much more general self-interaction potential
(\ref{eq:potential}).

It is worth mentioning that this mapping is only effective for
$a\not=\pm 1$, and for $a^2<1$ it preserves the unitarity of
both actions. Additionally, for $a=0$ it reduces to the
identity. Regarding the interpretation of the parameter $a$,
notice that if the starting conformal configuration vanishes at
infinity then this parameter in the scalar map
(\ref{eq:mapPhi}) is just related to the constant value
$\bar{\Phi}_0$ of the bar field at infinity, i.e.\
$a=\sqrt{\kappa\xi_D}\bar{\Phi}_0$. As stressed before, the
metric transformation (\ref{eq:mapg}) is easily understood as a
conformal transformation to a precise conformal frame. In
contrast, the meaning of the scalar transformation
(\ref{eq:mapPhi}) is more subtle, but for $a^2<1$ it may be
viewed as a $\mathrm{SL(2,\!I\!R)}$ transformation.

As an interesting remark that will be relevant in what follows,
one can mention that this mapping can be extended to a starting
action given by (\ref{eq:Sconformal}) supplemented with an
extra piece that is conformally invariant, for instance the
Maxwell action in $D=4$ or its nonlinear conformal extension in
arbitrary dimension \cite{Hassaine:2007py}. Indeed, under the
mapping (\ref{eq:map}) the additional action, being conformally
invariant, would remain in principle unchanged. Hence,
supplementing the conformal transformation on all the involved
fields with a trivial scaling in the additional field, the full
action will change by the same global factor to a bar action
given by (\ref{eq:Sbar}) together with the bar valued version
of the conformal extra piece. The procedure will become more
clear in the next section since we shall include the Maxwell
action in the applications of the mapping in $D=4$.

We would like to point out that for a free conformal scalar
field in the absence of cosmological constant, i.e.\
$\lambda=0$ and $\Lambda=0$, it is easy to check from both,
(\ref{eq:potential}) and (\ref{eq:barLambda}), that the
proposed transformation constitutes just a scaling of the
self-gravitating free conformal action to itself and
consequently a symmetry of the involved equations of motion.
Starting in this case from the Bekenstein black hole the
wormhole solution \cite{Barcelo:1999hq} is obtained, see
Ref.~\cite{Astorino:2014mda} for a recent discussion on this
point. In the next sections we concentrate in the cases where
$\lambda\ne0$ and $\Lambda\ne0$ and generate new classes of
solutions connected by the map to well-known conformal seeds.

\section{\label{sec:D=4} Generating new solutions in $D=4$}

We now proceed to exploit the method of generating solutions
for self-gravitating (super-)renormalizable scalar sources
conformally coupled to gravity, which are rigged by equations
(\ref{eq:FE}), from known conformal solutions of equations
(\ref{eq:NMGSconf1}). As stressed before, the map
(\ref{eq:map}) can be extended in order to include an extra
source that is also conformally invariant. We consider this
option in four dimensions with the Maxwell action
\begin{equation}\label{eq:SMax}
S_{\mathrm{M}}[g,A]=-\frac{1}{16\pi}\int{d^4x\,\sqrt{-g}\,F^{\mu\nu}F_{\mu\nu} },
\end{equation}
which is conformally invariant. Nevertheless, in order for the
Maxwell action to be mapped with the same global factor
appearing in (\ref{eq:mapaction}), the map (\ref{eq:map}) must
be extended with a trivial scaling only acting on the vector
potential as
\begin{equation}\label{eq:barA}
\bar{A}_{\mu}=\sqrt{1-a^2}A_{\mu}.
\end{equation}
In this case, the Maxwell action (\ref{eq:SMax}) together with
the self-gravitating conformal action (\ref{eq:Sconformal}) in
four dimensions are mapped, under the transformation
(\ref{eq:map}) supplemented by the scaling (\ref{eq:barA}), to
the general action
\begin{equation}\label{eq:D=4extmap}
\bar{S}[\bar{g},\bar{\Phi}]+S_{\mathrm{M}}[\bar{g},\bar{A}]
=(1-a^{2})\left(S[g,\Phi]+S_{\mathrm{M}}[g,A]\right).
\end{equation}
In four dimensions, there exists an asymptotically (A)dS
electrically charged black hole solution with nontrivial scalar
field obeying the field equations obtained from the variation
of the action $S[g,\Phi]+S_{\mathrm{M}}[g,A]$. This solution
corresponds to the charged version of the MTZ black hole
\cite{Martinez:2002ru} found in \cite{Martinez:2005di}, and
given by
\begin{subequations}\label{eq:MST}
\begin{align}
\bm{ds}^{2}={}&
 -\left[-\frac{\Lambda r^{2}}{3}+k\left(1-\frac{M}{r}\right)^{2}\right]\bm{dt}^{2}
\nonumber\\
&+\left[-\frac{\Lambda r^{2}}{3}+k\left(1-\frac{M}{r}\right)^{2}\right]^{-1}\bm{dr}^{2}
+r^{2}\bm{d\Omega}^{2}_{k},\label{eq:gMST}\\
\Phi(r)={}&\sqrt{-\frac{\Lambda}{6\lambda}}\,\frac{M}{r-M},\label{eq:phiMST}\\
\bm{A}(r)={}&-\frac{q}{r}\bm{dt},\label{eq:AMST}
\end{align}
where the two-dimensional base manifold is of constant
curvature denoted by $k$ and the constant $M$ related to the
mass together with the electric charge $q$ are tied via the
coupling constants as follows
\begin{equation}\label{eq:finetuneMST}
q^{2}=\frac{2\pi}{9}\frac{kM^{2}\left(\kappa\Lambda+36\lambda \right)}{\kappa\lambda}.
\end{equation}
\end{subequations}
In the neutral limit $q=0$ it becomes the original MTZ black
hole \cite{Martinez:2002ru} with its known fine tuning between
the coupling constants, $\Lambda/\lambda=-36/\kappa$. For
$\Lambda=0=\lambda$ and $k=1$ the MTZ black hole reduces to the
Bekenstein one \cite{Bekenstein:1974sf}. The solution
(\ref{eq:MST}) will be our conformal seed configuration in
order to generate solutions of the field equations obtained
from the variation of the action at the left hand side of
(\ref{eq:D=4extmap}), defined by
\begin{align}
\bar{G}_{\mu\nu}+\bar{\Lambda}\bar{{g}}_{\mu\nu}={}&{\kappa}\left[\bar{T}_{\mu\nu}+
\frac{1}{4\pi}\left(\bar{F}_{\mu\sigma}\bar{F}_{\nu}^{~\sigma}
-\frac{1}{4}\bar{g}_{\mu\nu}\bar{F}_{\alpha\beta}\bar{F}^{\alpha\beta}\right)\right],
\nonumber\\
\bar{\Box}\bar{\Phi}-\frac{1}{6}\bar{R}\bar{\Phi} ={}& \frac{dV(\bar{\Phi})}{d\bar{\Phi}},
\label{eqmotionM2}\\
\bar{\nabla}_{\mu}\bar{F}^{\mu\nu}={}&0,\nonumber
\end{align}
where the potential (\ref{eq:potential}) is given in four
dimensions by the (super-)renormalizable one
\begin{subequations}\label{eq:potentialD4}
\begin{equation}
V(\bar{\Phi})=\lambda_1\bar{\Phi}+\lambda_2\bar{\Phi}^2+\lambda_3\bar{\Phi}^3
+\lambda_4\bar{\Phi}^4,
\end{equation}
with couplings constants determined by
\begin{align}
\lambda_1&=-\frac{2\sqrt{6}}{3}\frac{a(\kappa\Lambda+36a^{2}\lambda)}
{\kappa^{3/2}(1-a^2)^{3}},\\
\lambda_2&= \frac{a^{2}(\kappa\Lambda+36\lambda)}{\kappa(1-a^{2})^{3}},
\label{eq:lambda2}\\
\lambda_3&=-\frac{\sqrt{6}}{9}\frac{a(a^{2}\kappa\Lambda+36\lambda)}
{\kappa^{1/2}(1-a^{2})^{3}},\\
\lambda_4&= \frac{1}{36}\frac{a^{4}\kappa\Lambda+36\lambda}{(1-a^{2})^{3}},
\label{eq:lambda4}
\end{align}
\end{subequations}
and the cosmological constant takes the following expression
\begin{equation}\label{eq:LambdaD4}
\bar{\Lambda}=\frac{\kappa\Lambda+36a^{4}\lambda}{\kappa(1-a^{2})^{3}}.
\end{equation}
The above parameterizations can be understood as follows: the
transformations (\ref{eq:LambdaD4}) and (\ref{eq:lambda4}) are
just a one-parameter invertible linear map between the initial
cosmological and renormalizable coupling constants
$(\Lambda,\lambda)$ and the final ones
$(\bar{\Lambda},\lambda_4)$. Rewriting the rest of the
parameterizations in terms of $(\bar{\Lambda},\lambda_4)$, via
this invertible linear map, one can consider the final
cosmological and renormalizable coupling constants as arbitrary
and the full transformations just describe a one-parameter
subspace of the three-dimensional parameter space
$(\lambda_1,\lambda_2,\lambda_3)$ characterizing the strictly
super-renormalizable contributions. This is the precise
subspace of the general problem with (super-)renormalizable
self-interactions which is accessible via the mapping
(\ref{eq:map}) from the conformal sector.

The map acting on (\ref{eq:MST}) gives rise to a new charged
solution of the field equations (\ref{eqmotionM2}) that reads
\begin{subequations}\label{eq:mapMST}
\begin{align}
\bm{d\bar{s}}^{2}={}&
\left(\frac{r-M\left(1-\sqrt{-\frac{\kappa\Lambda}{36\lambda} }a\right)}{r-M}\right)^{2}
\nonumber\\
&\times\Bigg\{
 -\left[\frac{-\Lambda r^{2}}{3}+k\left(1-\frac{M}{r}\right)^{2}\right]\bm{dt}^{2}
\nonumber\\
&\qquad+\left[\frac{-\Lambda r^{2}}{3}+k\left(1-\frac{M}{r}\right)^{2}\right]^{-1}\bm{dr}^{2}
+r^{2}\bm{d\Omega}^{2}_{k}\Bigg\}, \label{eq:transformedgMST}\\
\bar{\Phi}(r)={}&\sqrt{\frac{6}{\kappa}}\,
\frac{ar+M\left(\sqrt{-\frac{\kappa\Lambda}{36\lambda}}-a\right)}
{r-M\left(1-\sqrt{-\frac{\kappa\Lambda}{36\lambda}}a\right)},
\label{eq:transformedphiMTZ}\\
\bar{\bm{A}}(r)={}&-\frac{\bar{q}}{r}\bm{dt},
\end{align}
where the charge $\bar{q}$ is also tied in terms of the
integration constant $M$ via the original coupling constants as
\begin{equation}\label{eq:barq}
\bar{q}^{2}=(1-a^{2})\frac{2{\pi}}{9}\frac{kM^{2}
\left(\kappa\Lambda+36\lambda\right)}{\kappa\lambda}.
\end{equation}
\end{subequations}
We can easily check that in the neutral limit $\bar{q}=0$, the
relation (\ref{eq:barq}) leads to the same fine tuning between
the original coupling constants characterizing the MTZ black
hole,
\begin{equation}\label{eq:finetuneMTZ}
\lambda=-\frac{1}{36}\kappa\Lambda,
\end{equation}
and the solution becomes the Anabalon-Cisterna solution
\cite{Anabalon:2012tu} which is supported by a potential
involving all the (super-)renormalizable terms except the
massive one. This can easily be explained through our mapping
since for this specific fine tuning (\ref{eq:finetuneMTZ}), the
coupling constant defining the mass (\ref{eq:lambda2}) vanishes
identically. In fact, the AC solution \cite{Anabalon:2012tu} is
exactly the result of applying the proposed map to the MTZ
black hole \cite{Martinez:2002ru}, that's why it must respect
the MTZ fine tuning (\ref{eq:finetuneMTZ}). This situation
changes when the electric charge is introduced since the
involved fine tuning is now understood as a relation between
the integration constants and not as one between the coupling
constants, which in turns allows the mass term to remain turned
on. The AC solution represents black holes and wormholes, and
this is also the case in the above charged generalization
(\ref{eq:mapMST}).

In the next section, we will pursue this strategy in the
three-dimensional case where there exist more conformal seed
configurations.

\section{\label{sec:D=2+1}Generating new solutions in $D=2+1$}

In three dimensions, we have a priori more configurations that
are interesting solutions of the Einstein equations supported
by a conformal scalar field (\ref{eq:NMGSconf1}). On the one
hand there is the well-known Mart\'{\i}nez-Zanelli conformal
black hole \cite{Martinez:1996gn}, its self-interacting
generalization \cite{Henneaux:2002wm} and a time-dependent
solution describing the exact gravitational collapse to the
last black hole \cite{Xu:2014xqa}. On the other hand there also
exist a different kind of time-dependent solutions named
stealth configurations characterized by the peculiarity that
both sides (the gravity and the matter source part) of Einstein
equations vanish independently. Indeed, in three dimensions, it
has been shown in \cite{AyonBeato:2004ig} the existence of a
time-dependent nontrivial self-interacting scalar field
nonminimally coupled to gravity having an energy-momentum
tensor that vanishes on the BTZ black hole
\cite{Banados:1992wn}. Finally, there also exist conformal
solutions supporting AdS-waves \cite{AyonBeato:2005qq}.

In this section we will provide more new examples of
self-gravitating scalar field solutions with
(super-)renormalizable self-interaction starting from almost
all the previously mentioned conformal seeds via the mapping
(\ref{eq:map}). We start by specifying the
(super-)renormalizable potential (\ref{eq:potential}) resulting
from the map at $D=3$, which will be the common support for all
the generated solutions
\begin{subequations}\label{eq:potentialD3}
\begin{equation}
V(\bar{\Phi})=\lambda_1\bar{\Phi}+\lambda_2\bar{\Phi}^2+\lambda_3\bar{\Phi}^3
+\lambda_4\bar{\Phi}^4+\lambda_5\bar{\Phi}^5+\lambda_6\bar{\Phi}^6,
\end{equation}
where the coupling constants are fixed as
\begin{align}
\lambda_{1}&=-\frac{3}{\sqrt{2}}\frac{a(\kappa^{2}\Lambda+512a^{4}\lambda)}
{\kappa^{5/2}(1-a^2)^5},\\
\lambda_{2}&=\frac{15}{8}\frac{a^{2}(\kappa^{2}\Lambda+512a^{2}\lambda )}
{\kappa^{2}(1-a^2)^5},\\
\lambda_{3}&=-\frac{5\sqrt{2}}{8}\frac{  a^{3}(\kappa^{2}\Lambda+512\lambda)}
{\kappa^{3/2}(1- a^2)^5},\\
\lambda_{4}&=\frac{15}{64}\frac{a^{2}(a^{2}\kappa^{2}\Lambda+512\lambda)}{\kappa(1-a^2)^5},\\
\lambda_{5}&=-\frac{3\sqrt{2}}{128}\frac{a(a^{4}\kappa^{2}\Lambda+512\lambda)}
{\kappa^{1/2}(1-a^2)^5},\\
\lambda_{6}&=\frac{1}{512}\frac{a^{6}\kappa^{2}\Lambda+512\lambda}{(1-a^2)^5},
\label{eq:lambda6}
\end{align}
\end{subequations}
along with the relationship between the new and the original
cosmological constants,
\begin{equation}
\label{eq:LambdaD3}
\bar{\Lambda}=\frac{\kappa^{2}\Lambda+512a^{6}\lambda}{\kappa^{2}(1-a^{2})^5}.
\end{equation}
As in the four-dimensional case, the above parameterizations
have the following interpretation: the transformations
(\ref{eq:LambdaD3}) and (\ref{eq:lambda6}) are again a
one-parameter invertible linear map between the initial
cosmological and renormalizable coupling constants
$(\Lambda,\lambda)$ and the final ones
$(\bar{\Lambda},\lambda_6)$. This allows to rewrite the rest of
the parameterizations in terms of $(\bar{\Lambda},\lambda_6)$
and finally consider them as arbitrary. Hence, the full
transformations again describe a one-parameter subspace of the
five-dimensional parameter space $(\lambda_1,\ldots,\lambda_5)$
defined by the strictly super-renormalizable contributions.
This is the only subspace of the general problem with
(super-)renormalizable self-interactions than can be probed in
three dimensions from the conformal sector using the mapping
(\ref{eq:map}).

Let us first consider the conformal seed configuration which
corresponds to the self-interacting version of the MZ conformal
black hole \cite{Martinez:1996gn} found in
\cite{Henneaux:2002wm}, which is a solution of the field
equations (\ref{eq:NMGSconf1}),
\begin{subequations}\label{eq:gintMZ}
\begin{align}
\bm{ds^{2}}={}&-\bigg(\frac{r^2}{l^2}-\frac{2\alpha B^3}{r}-3\alpha B^2\bigg)\bm{dt^2}
\nonumber\\
&+\bigg(\frac{r^2}{l^2}-\frac{2\alpha B^3}{r}-3\alpha B^2\bigg)^{-1}\bm{dr^2}
+r^{2}\bm{d\varphi^2},\label{eq:gMZ}\\
\Phi(r)={}&\sqrt{\frac{8B}{\kappa(r+B)}},\label{eq:phiMZ}\\
\alpha\equiv{}&\frac{\kappa^{2}-512\lambda l^{2}}{\kappa^{2}l^{2}},
\label{eq:alpha}
\end{align}
\end{subequations}
where the cosmological constant is chosen to be negative
$\Lambda=-1/l^{2}$ and $B$ is an integration constant. Hence,
as in the four-dimensional case, a solution of the field
equations (\ref{eq:FE}) with the (super-)renormalizable
potential given by (\ref{eq:potentialD3}) can be constructed
from the conformal seed (\ref{eq:gintMZ}), and the resulting
configuration is given by
\begin{subequations}\label{eq:transformedMZ}
\begin{align}
\label{eq:transformedgMZ}
\bm{d\bar{s}}^{2}={}&\left(\frac{a\sqrt{B}+\sqrt{r+B}}{\sqrt{r+B}}\right)^4\nonumber\\
&\times\Bigg[
 -\bigg(\frac{r^{2}}{l^{2}}-\frac{2\alpha B^{3}}{r}-3\alpha B^{2}\bigg)\bm{dt}^{2}\nonumber\\
&\qquad+\bigg(\frac{r^{2}}{l^{2}}-\frac{2\alpha B^{3}}{r}-3\alpha B^{2}\bigg)^{-1}\bm{dr}^{2}
+r^{2}\bm{d\varphi}^{2}\Bigg],\\
\bar{\Phi}(r)={}&\sqrt{\frac{8}{\kappa}}\frac{\sqrt{B}+a\sqrt{r+B}}{a\sqrt{B}+\sqrt{r+B}}.
\label{eq:transformedphiMZ}
\end{align}
\end{subequations}
We remark that, as in the four-dimensional case, one could have
started from the charged version of the solution
(\ref{eq:gintMZ}) given in \cite{Cardenas:2014kaa}, which is
supported by a conformally invariant nonlinear electrodynamics
\cite{Hassaine:2007py}. Hence, in this case the map will yield
to a nonlinearly charged version of the configuration
(\ref{eq:transformedMZ}) based on the same conformal
electrodynamics, but where the scalar field self-interact with
the non-conformal (super-)renormalizable potential
(\ref{eq:potentialD3}).

As a second conformal seed, we now consider the exact
gravitational collapse of the self-interacting version of the
MZ conformal black hole (\ref{eq:gintMZ}) derived in
\cite{Xu:2014xqa}. This is a time-dependent solution which in
the limit when time goes to infinity reduces to the
configuration (\ref{eq:gintMZ}). We stress that it is one of
the few examples of an exact gravitational collapse in the
literature. Due to time dependence it is more convenient to
write the metric solution in the Eddington-Finkelstein
coordinates as
\begin{subequations}\label{eq:Xu}
\begin{align}
\bm{ds}^{2}={}&-f(u)^{-2/3}\bigg( \frac{r^{2}}{l^{2}}-\frac{2\alpha B^{3}}{r}f(u)^{2}
\nonumber\\
&-3\alpha B^{2}f(u)^{2/3} \bigg)\bm{du}^{2}+2f(u)^{-1/3}\bm{dudr}
\nonumber\\
&+r^{2}\bm{d\varphi}^{2},\\
\Phi(u,r)={}&\sqrt{ \frac{8B}{\kappa\left( rf(u)^{-4/3} +B\right)}},\\
f(u)\equiv{}&\tanh\left( \frac{3\alpha B u}{8}\right),
\end{align}
\end{subequations}
where $B$ is an integration constant and $\alpha$ is defined in
terms of the coupling constants in (\ref{eq:alpha}). Notice
that in the limit of $u\rightarrow\infty $ we have $f(u)=1$,
i.e.\ the black hole solution (\ref{eq:gintMZ}) is just the
final state of the evolutive solution (\ref{eq:Xu}). As before,
starting from this conformal seed solution, one can generate a
new time-dependent solution of the field equations
(\ref{eq:FE}) with a (super-)renormalizable potential fixed by
(\ref{eq:potentialD3})
\begin{subequations}\label{eq:transformedgCollapse}
\begin{align}
\bm{d\bar{s}}^{2}={}&\left(a\sqrt{\frac{B}{rf(u)^{-4/3}+B }}+1\right)^{4}
\nonumber\\
&\times\Bigg[-f(u)^{-2/3}\bigg( \frac{r^{2}}{l^{2}}-\frac{2\alpha B^{3}}{r}f(u)^{2}
\nonumber\\
&\qquad-3\alpha B^{2}f(u)^{2/3}\bigg)\bm{du}^{2}+2f(u)^{-1/3}\bm{dudr}
\nonumber\\
&\qquad+r^{2}\bm{d\varphi}^{2}\Bigg],\\
\label{eq:transformedphiCollapse}
\bar{\Phi}(u,r)={}&\sqrt{\frac{8}{\kappa}}\,\frac{\sqrt{B}+a\sqrt{rf(u)^{-4/3}+B}}
{a\sqrt{B}+\sqrt{rf(u)^{-4/3}+B}}.
\end{align}
\end{subequations}
Consequently, the final state of this evolution $u\to\infty$ is just the
previously generated stationary solution
(\ref{eq:transformedMZ}).

As the last conformal seed, we consider again a time-dependent
configuration but of a different sort. It represents a
conformal stealth \cite{AyonBeato:2004ig} overflying the BTZ
black hole \cite{Banados:1992wn}. The stealths are particular
nontrivial solutions of Einstein equations where both of its
sides vanish independently, that is
\begin{equation}
G_{\mu\nu}-\frac{1}{l^2}g_{\mu\nu}=0=\kappa T_{\mu\nu}.
\label{stealtheq}
\end{equation}
The left hand side produces the BTZ black hole
\cite{Banados:1992wn} assuming only rotational symmetry
\cite{AyonBeato:2004if}, while the right hand side has
nontrivial solutions only when this black hole is static
\cite{AyonBeato:2004ig}. The resulting conformal stealth is
given in the Eddington-Finkelstein coordinates by
\begin{subequations}
\begin{align}
\label{eq:BTZ}
\bm{ds}^{2}={}&-\left(\frac{r^{2}}{l^{2}}-M\right)\bm{du}^{2}
-2\bm{du dr}+r^{2}\bm{d\varphi}^{2},\\
\label{eq:phiStealth}
\Phi(u,r)={}&\sqrt{\frac{8}{\kappa}}\frac{1}{\sqrt{\sigma(u,r)}},\\
\sigma(u,r)\equiv{}&
\sqrt{\frac{8l^2\lambda-h^2}{Ml^{2}}}
\Bigg[ r\cosh\left(\frac{\sqrt{M}u}{l}\right)\nonumber\\
&+\sqrt{M}l\sinh\left( \frac{\sqrt{M}u}{l}\right)\Bigg]+h.
\end{align}
\end{subequations}
It is interesting to note that the mapped solution given by
\begin{subequations}
\begin{align}
\label{transformedBTZ}
\bm{d\bar{s}}^{2}={}&\left(\frac{a}{\sqrt{\sigma(u,r)}}+1\right)^4
\bigg[-\left(\frac{r^{2}}{l^{2}}-M\right)\bm{du}^{2}\\
&\qquad\qquad\qquad\qquad\quad-2\bm{dudr}+r^{2}\bm{d\varphi}^{2} \bigg],
\nonumber\\
\label{eq:transformedphiStealth}
\bar{\Phi}(u,r)={}&\sqrt{\frac{8}{\kappa}}\,
\frac{1+a\sqrt{\sigma(u,r)}}{a+\sqrt{\sigma(u,r)}},
\end{align}
\end{subequations}
is as usual a new solution of the field equations (\ref{eq:FE})
exhibiting also time dependence, but now it is not longer a
stealth configuration (\ref{stealtheq}) for the bar fields.

\section{\label{sec:conclu}Conclusions}

Here, we provide new examples of self-gravitating solutions in
presence of a cosmological constant $\bar{\Lambda}$ for scalar
fields conformally coupled to gravity, allowed to self-interact
with themselves via a potential where all the
super-renormalizable contributions, defined by the positive
mass-dimension coupling constants
$(\lambda_1,\ldots,\lambda_{(D+2)/(D-2)})$, and the
renormalizable one described by the dimensionless coupling
constant $\lambda_{2D/(D-2)}$ are turned on. This was possible
due to the introduction of a new one-parameter mapping
connecting any self-gravitating conformal scalar solution with
the previous configurations. The map consists in a conformal
transformation for the metric and a $\mathrm{SL(2,\!I\!R)}$
transformation for the scalar field. All the studied cases
allow the following interpretation of the mapping: the sector
of the general problem with (super-)renormalizable
self-interactions that can be probed with a conformal
counterpart, is the one where the cosmological and
renormalizable coupling constants, $\bar{\Lambda}$ and
$\lambda_{2D/(D-2)}$, are arbitrary but from the
$(D+2)/(D-2)$-dimensional parameter space
$(\lambda_1,\ldots,\lambda_{(D+2)/(D-2)})$ of the
super-renormalizable contributions only a one-parameter
subspace described by the mapping is accessible.

We systematically use well-known conformal solutions in the
literature as seed configurations in this map to generate the
new (super-)renormalizably dressed solutions. Concretely, we
exhibit a charged version of the AC solution at $D=4$
\cite{Anabalon:2012tu}, starting from the charged version
\cite{Martinez:2005di} of the MTZ black hole
\cite{Martinez:2002ru}. We explain why the AC configurations
have no mass term in the potential and how this mass term
appears now due to the presence of the electric charge. Many
other similar examples are generated in $D=3$. The first of
them is generated from the generalization of the MZ black hole
\cite{Martinez:1996gn} including a conformal self-interaction
\cite{Henneaux:2002wm}. A second one is generated from the
exact gravitational collapse of the previous conformal black
hole \cite{Xu:2014xqa}, giving this time a
(super-)renormalizably dressed time-dependent configuration. As
a last example, the conformal stealth overflying the static BTZ
black hole is used as seed \cite{AyonBeato:2004ig}. The
resulting configuration is again time dependent but, since the
map mix the gravity and matter contributions, the resulting
solution is not longer a stealth configuration.

All the generated (super-)renormalizably dressed configurations
have the same asymptotic as their conformal seeds since the
involved conformal factors take unit value at infinity, i.e.\
they are all asymptotically (A)dS spacetimes. On the contrary,
the (super-)renormalizably self-interacting scalar fields no
longer vanish at infinity as its conformal counterparts. In
fact, the parameter of the introduced mapping defines the
constant value of the final fields at infinity. However, it
cannot be interpreted as a new integration constant since as
previously discussed this parameter appears in the action
probing the strictly super-renormalizable sector.

Finally, at dimension $D=6$ our generating technic can also
provide (super-)renormalizably dressed solutions, this time for
potentials built from cubic polynomials. However, the only
known conformal seeds in this dimension to our knowledge are
those consisting of stealth configurations and AdS-waves
\cite{AyonBeato:2006jf}. The relevant stealth examples live on
flat spacetime \cite{AyonBeato:2005tu} and (A)dS$_6$ spacetime
\cite{Ayon-Beato:SAdS}. We leave the exploration of the
resulting six-dimensional configurations for future work.


\begin{acknowledgments}
We thank the organizers of the fourth GravUach conference at
Valdivia as well as the participants for useful discussions,
specially to A.~Anabalon and A.~Cisterna. EAB is partially
supported by grants 175993 and 178346 from CONACyT, grants
1121031, 1130423 and 1141073 from FONDECYT and ``Programa
Atracci\'{o}n de Capital Humano Avanzado del Extranjero, MEC''
from CONICYT. MH is partially supported by grant 1130423 from
FONDECYT. JAMZ is supported by grant 243377 and ``Programa de Becas
Mixtas'' both from CONACyT.
\end{acknowledgments}


\end{document}